# Gross beta determination using a scintillating fiber array detector for drinking water

Wen-hui Lv[a,b], Hongchang Yi[a], Tong-Qing Liu[a], Zhi Zeng[a1], Jun-li Li[a], Hui Zhang[a], Hao Ma[a]

[a] Key Laboratory of Particle and Radiation Imaging (Ministry of Education) and Department of Engineering Physics, Tsinghua University, Beijing 100084, China
[b] Xi'an Research Institute of Hi-Tech, Xi'an 710025, China

**Abstract:** A scintillating fiber array measurement system for gross beta is developed to achieve real-time monitoring of radioactivity in drinking water. The detector consists of 1,096 scintillating fibers, both sides of the fibers connect to a photomultiplier tube, and they are placed in a stainless steel tank. The detector parameters of working voltage, background counting rate and stability of the detector were tested, and the detection efficiency was calibrated using a standard solution of potassium chloride. Through experiment, the background counting rate of the detector is 38.31 cps and the detection efficiency for β particles is 0.37 cps/(Bq/l); the detector can reach its detection limit of 1.0 Bq/l for β particles within 100 minutes without pre-concentration.

## I Introduction

Radioactive safety of drinking water is very important in many countries. Since the concentration of radionuclides in drinking water is very low in normally, the determination of concentration of each radionuclides would be a time-consuming and costly work. To avoid the procedure and save time, the total radioactivity of gross alpha and beta in drinking water are usually used to screening the samples at first. If the total radioactivity is higher than a screening level, the radionuclide analysis would be implemented in the following [1].

The recommendation of screening level of total radioactive in the drinking water are mainly suggested by World Health Organization (WHO), the European Union (EU) and the Unite State Environmental Protection Agency (USEPA) [2-4]. The recommendation screening levels of gross alpha/beta radioactivity in drinking water by WHO is 0.5 Bq/l and 1.0 Bq/l , while by EC is 0.1 Bq/l and 1.0 Bq/l, and for USEPA the screening levels are 0.19 Bq/l for gross alpha activity and 0.56 Bq/l for gross beta activity[5].

The measurement technology of gross alpha and beta in drinking water includes the evaporation concentration method, the co-precipitation method, and real-time monitoring. The evaporation concentration method usually takes a 0.1–2.0 L sample as total dissolved solid, and the sample is evaporated slowly to complete dryness, then the gross alpha and beta radioactivity levels can be calculated by measuring those of the solid residues[6,7], and the detector can be a proportional counter[8,9] or scintillator detector[10,11]. This method has the advantages of the reliable measurement results and lower detection limit, but the sampling period is long and the sample process is complex, making it rather inefficient, especially in the case of an alert for unexpected radioactive events[12,13], and this method does not allow for the determination of volatile radionuclides, making the results underestimated. The co-precipitation method sets the pH value of the filtered water sample and heats it to purge radon and $CO_2$. Then, the radium isotopes are co-precipitated with $Ba^{2+}$ as $Ba(Ra)SO_4$, whereas uranium, thorium, and polonium isotopes can be co-precipitated with $Fe(OH)_3$ by adding a $Fe^{3+}$ carrier, and finally $NH_4OH$ is used to adjust the pH value. The gross alpha and beta activity of the filtered and dried precipitate is counted by a detector, which can be a ZnS(Ag) scintillator detector[14,15]. This method needs more chemical treatment than the evaporation methods. Real time monitoring measures the water sample directly and gives the gross alpha and beta radioactivity levels in real time. This technology is currently in the research stage.

---
[1] Corresponding author: Zhi Zeng, email address: zengzhi@tsinghua.edu.cn

The European Union are developing a tap water radioactivity–real time monitoring (TAWARA_RTM) system[16] that uses a large area 200 × 200 $mm^2$ EJ-444 scintillation detector to monitor gross alpha and beta activity in tap water. The possibility of stacking many detectors and thus getting a compact device with total active area of 1.0 $m^2$. The system is conceived to detect gross alpha/beta activity in the order 1.0 Bq/l in several tens of minutes. The instrument parameters of sensitivity, selectivity, background, short/long term stability, linearity with respect to activity are tested[17], and the calibration work for the TAWARA system is done with standard solution of $^{241}$Am for gross alpha and $^{60}$Co, $^{40}$K, $^{90}$Sr, $^{90}$Y, $^{18}$F for gross beta.

This research develops a gross beta radioactivity monitoring system with scintillating fiber array. The system will provide a method of real-time monitoring for gross beta activity in drinking water to verify whether the water is well within the limits or it is reach the threshold that requires blocking the distribution to the public.

The scintillating fiber array measurement system consists of scintillating fiber detector, electronic module, and data processing and acquisition module. The scintillating fiber is used as both detection medium and fluorescence transmission medium. Through design, it can achieve the detection of gross beta activity in the order of 1.0 Bq/l in 100 minutes.

## II Framework of the system

The scintillating fiber array measurement system is composed of a detection module, an electronic module, and a data processing and acquisition module; Figure 1 is a structural schematic diagram of detector.

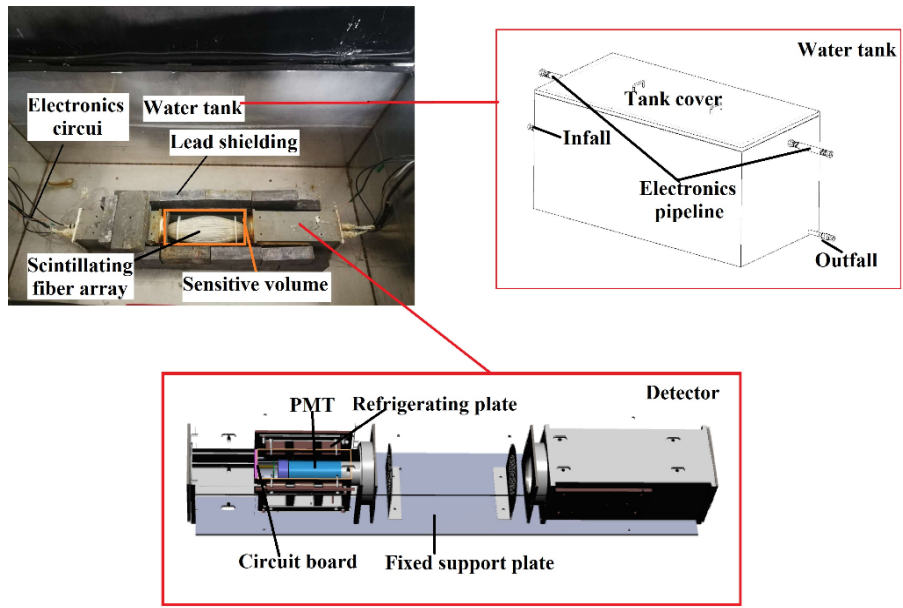

Fig. 1 Structure schematic diagram of the system

The detection system consists of 1,096 scintillating fibers which type is BCF-10, the diameter of each fiber is 1 mm, the length is 40 cm, and the distance between two scintillating fibers is 2.5 mm. Both ends of the fibers are fastened into bundles and connected to photomultiplier tubes, and the cross sections of the fibers are polished to ensure smoothness. The photomultiplier tubes are sealed in polyvinyl chloride (PVC), the scintillating fiber and PVC shell are pressed by screw and sealed with silicone rubber to stop water entering into the PVC. The detector is placed in a rectangular stainless steel water tank of volume $160 \times 80 \times 70$ cm$^3$, and the water tank is sealed to exclude any light. The electronic module is composed of power supply and signal processing circuit, which the power supply can provide power for detector and electronic circuits, and the processing circuit deals with the current signals. Data processing and acquisition module receives the current signal and converts it to counts. This module has the function of adjusting the sampling time and voltage, automatic measurement and storage. Finally, the detection system gives gross beta activity and provides a threshold alarm signal.

Adding the solution in the water tank so that the scintillating fibers and solution are fully mixed. Radiation particles in the sensitive volume deposit their energies into the fibers and convert to fluorescence, fluorescence transmits to both ends of the fibers by total reflection and produce photoelectrons on the cathode of the photomultiplier tube. After pulse shaping, baseline recovery, anti-accumulation circuit and pulse amplitude discrimination, the signals are recorded by a counter. The results can be total counts for β particles and gross β activity concentration, the relationship between the total counts and gross β activity concentration is established through detection efficiency calibration. The scintillating fiber detector is capable of measuring β particles with an energy greater than 40 keV.

To ensure the stability of the instrument for a long period of time, the scintillating fiber array detector should be cleaned using tap water and purified water respectively after the sample has been measured. First, the detector is rinsed by tap water for half an hour, then with purified water for ten minutes. If the detector needs continuous monitoring, decontamination and cleaning of the detector are required every month.

# III Experiment and Results

## 3.1 Voltage test without water

The working voltage of the photomultiplier tube has a great influence on the photocurrent, usually it is less than 1,000 V, otherwise it will cause self-discharge to damage the tube. The optimal voltage point is determined according to the characteristics of plateau curve of photomultiplier tube with potassium chloride(KCl) crystal at voltage of 500~1,100 V. Weighed 100g KCl crystal and placed it in a thin sealed bag. Put the bag under the fiber and flattened it, the bag was 2mm away from the scintillating fibers. The detector was surrounded by lead bricks to shield the environment radiation, and the water tank was closed and pressed to prevent visible light. The measurement time was 60 s, and measured 120 times. Figure 2 was the average counts for KCl crystal for 120 times with different voltages.

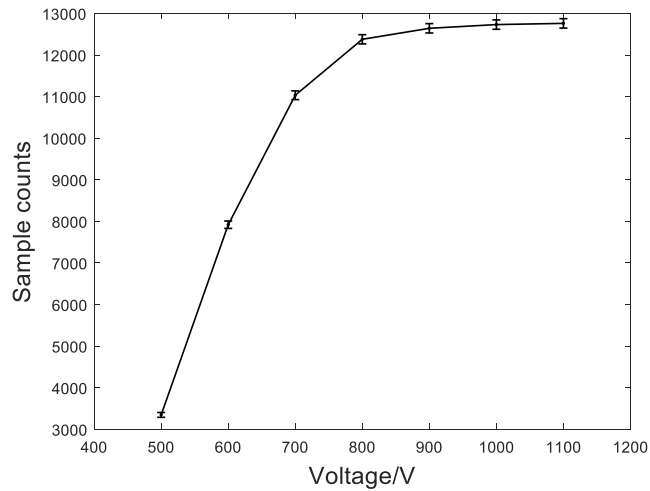

Fig. 2 Plateau curve of the photomultiplier tube

The plateau curve of the photomultiplier tube showed that the plateau region was 800 V~1,100 V, to reduce the background and protect the photomultiplier tube, the voltage point was selected 800 V.

## 3.2 Background test with water

For low-level counting devices the background counting rate must satisfy a Poisson distribution. The stability and distribution of background count of detector was tested.

Added 35cm deep water into the tank to make sure the detector was completely covered by the water. Lead bricks were placed around the detector to shield environment radiation, then the water tank was closed with screws. Adjusted the detector measurement parameters, the working voltage was 800 V and the measurement time was 600 s, and measured 120 times. The average counting rate was 38.31 ± 0.26 s$^{-1}$. The Poisson test method was as follows[18]:

(1) The background counts were divided into 10 groups, each group was measured 10 times, and each time measured 600 s.

(2) The average value and standard deviation of the background counts were calculated for each group.

(3) The calculated statistic $\chi^2$ for each group was:
$$\chi^2 = (n-1)S^2/N$$

(4) Comparing the $\chi^2$ value to the upper quantile of the $\chi^2$ distribution with α

conspicuous level, if the $\chi^2$ value was less than the upper quantile, the background counts were normal at a 1-α confidence level, otherwise there was a reason to doubt that the device was working correctly. The measurement results were listed in Table 1. Selecting the conspicuous level of α = 0.05, the upper quantile of the $\chi^2$ test was 16.92.

Table 1 $\chi^2$ test for background

| Group | Average counts | Standard deviation value | $\chi^2$ test value | Average counting rate |
|---|---|---|---|---|
| 1 | 22960.5 | 138.19 | 7.49 | 38.27 |
| 2 | 22909.7 | 144.02 | 8.15 | 38.18 |
| 3 | 22791.5 | 75.90 | 2.27 | 37.99 |
| 4 | 22909.7 | 104.95 | 4.33 | 38.18 |
| 5 | 22960.5 | 164.06 | 10.55 | 38.27 |
| 6 | 22927.9 | 165.77 | 10.79 | 38.21 |
| 7 | 22880.6 | 109.16 | 4.69 | 38.13 |
| 8 | 22873.5 | 176.77 | 12.30 | 38.12 |
| 9 | 22855.5 | 175.00 | 12.06 | 38.09 |
| 10 | 22955 | 100.01 | 3.92 | 38.26 |

All experimental results of $\chi^2$ test value were less than 16.92, and there was no abnormality in the background.

**3.3 Long term stability test**

The long term stability of background count of detector was tested. The experiments were divided into A and B groups, and both groups were adding 35cm deep water. The background count for group A was that the detector did not measure any radioactivity solution, while the background of group B was that the detector had been used in measuring KCl solution and after measurement it had been cleaned. The measurement conditions were the same with the background test. The measurement time was 600s, and measured 60 times(10h) for each group. The counting statistics distribution was shown in figure 3.

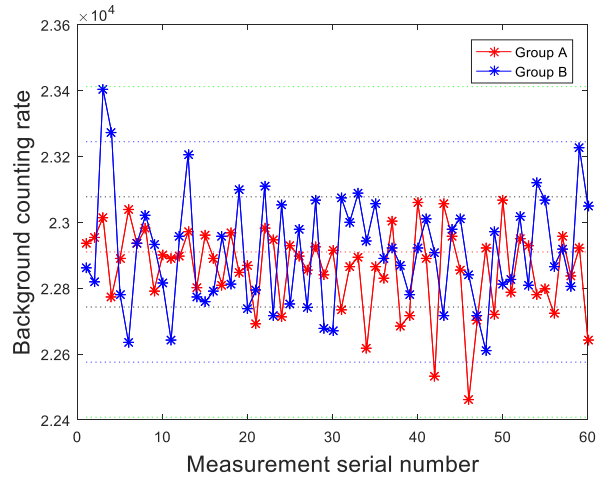
Fig. 3 Counting statistics distribution for different conditions

The statistics for long term repeatability of background was list in table 2. In table C stood for average count, u(C) was experimental standard deviation, $C^{1/2}$ was expected standard deviation under Poisson statistic assumption, R(C) was average counting rate, and u(R) was experimental standard deviation of average counting rate.

Table 2 Statistic for long-term repeatability of background

| Group | A | B |
| --- | --- | --- |
| C | 22859.7 | 22910.1 |
| u(C) | 126.05 | 167.21 |
| $C^{1/2}$ | 151.19 | 151.36 |
| R(C) | 38.10 | 38.18 |
| U(R) | 0.21 | 0.28 |

It is noted that the background counting rates before and after adding KCl solution were nearly the same and the error was 0.21%. The difference may be that:

a) The number of measurements is small, and the background count rate is within the statistical fluctuation.

b) The radioactivity concentration of KCl solution reaches 80 Bq/l, and the detector cleaning time is not enough.

### 3.4 Detection efficiency calibration

The detection efficiency of the scintillating fiber detector was tested using KCl solution with different activity. KCl crystal was weighed by balance of 0.1 mg accuracy, and it was configured as solutions with activity levels of 0.5 Bq/l, 1.0 Bq/l, 2.5 Bq/l, 6.0 Bq/l, 10.5 Bq/l, 20.0 Bq/l, 30.0 Bq/l, 50.0 Bq/l and 80.0 Bq/l. After the radioactive solution mixing uniformity, the cleaned lead bricks were placed around the detector.

The measurement time was 600 s, measured 120 times, and measured conditions was the same with above. The counts of detector versus activity of KCl solution were shown in Figure 4.

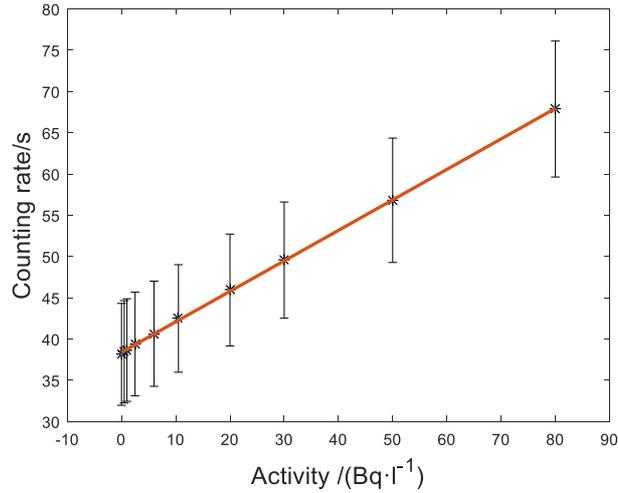

Fig. 4 Counting rate for KCl solutions with different activity

The relationship between the counting rate($R(C)$) of the detector and the radioactivity($C$) of the solution can be obtained through linear fitting.

$$R(C) = 0.37C + 38.37$$

The fitting result indicated that the detection efficiency of the detector was 0.37 cps/(Bq/l), and the background counting rate was 38.37 cps, which differed from the experimental measurement by 1%.

The standard KCl solution is used for detection efficiency calibration in this paper, When $^{40}$K emits beta particles, it also releases accompanying gamma rays. The energy deposition of γ rays for a single scintillating fiber can be neglected; however, the increase in scintillating fibers means that the γ rays will have an effect on the β measurements. Through Geant4 simulation, detector counts for γ rays generated by $^{40}$K in scintillating fiber accounted for 13% of the total counts. The influence and deduction of gamma rays will be further studied in the following work.

**3.5 Minimum detectable activity concentration**

In the field of radionuclide verification and environmental radioactivity monitoring, it is necessary to provide the minimum detectable activity concentration of radionuclides when the radioactivity level is lower than the background statistical fluctuation[19].

$$MDAC = \frac{2.71 + 4.65\sqrt{N_B}}{t\varepsilon P}$$

Taking the measurement time of 30 min as an example, the detection efficiency of the detector is 0.37 cps/(Bq/l), the background counting rate is 38.31 cps, and the branching ratio P is 1. Then, take the above parameters into the formula of MDAC; the MDAC value of the detector for gross beta is 1.84 Bq/l for 30min. If the MDAC reaches 1.0 Bq/l, the measurement time is 100 minutes.

Reducing the MDAC value or measurement time requires improving the detection efficiency and reducing the background. Joining the anti-coincidence detector can shield the cosmic ray and designing the structure of the scintillating fiber detector can improve the detection efficiency.

## IV Conclusion

The basic detection module for gross beta measurement system based on 1096

scintillating fiber detector was developed. The plateau curve of the detector was test, and the basic properties of background counts, long term stability, and detection efficiency were carried out, finally the MDAC of the detector was estimated. The background counting rate was 38.31 cps and the detection efficiency was 0.37 cps/(Bq/l). The detector can reach gross beta activity in the order of 1 Bq/l in 100 min. The scintillating fiber array detector can solve the problem of the real-time measurement of gross beta for drinking water, and it has engineering application value for the assessment of radioactive pollution and super threshold alarms.

**Acknowledgments**

This work was supported by National Key Scientific Instrument and Equipment Development Project (2016YFF0103902); Public science and technology research funds projects of ocean (No. 201505005).